\documentclass[12pt]{article}
\def \be{\begin{equation}}
\def \ee{\end{equation}}
\def \bea{\begin{eqnarray}}
\def \eea{\end{eqnarray}}
\def \p{\partial}
\def \nn{\nonumber}

\def \g{\gamma}

\def \th{\theta}
\def \G{\Gamma}
\def \o{\omega}
\def \O{\Omega}

\begin{document}

\title{The Time-dependent Supersymmetric Configurations in M-theory and Matrix Models}
\author{Bin Chen\footnote{email:bchen01@pku.edu.cn}\\ \\
Department of Physics, \\
Peking University\\
Beijing 100871, P.R. China}
 \maketitle

\begin{abstract}

In this paper, we study the half-supersymmetric time-dependent
configurations in M-theory and their matrix models. We find a
large class of 11D supergravity solutions, which keeps sixteen
supersymmetries. Furthermore, we investigate the isometries of
these configurations and show that in general these configurations
have no supernumerary supersymmetry. And also we define the matrix
models in these backgrounds following Discrete Light-Cone
Quantization (DLCQ) prescription.

\end{abstract}

\newpage

\section{Introduction}

It is very important in string theory to understand the mysterious
cosmological singularity. From string theory's point of view, such
singularity should be resolved by the stringy effects. It is
widely believed that near the big-bang/crunch singularity the
usual concept of spacetime in General Relativity would not be
applicable and there should be some new concepts and ideas to save
the life. Very recently several proposals have been raised to
address the issue\cite{Verlinde, Silverstein}.

One of the remarkable proposal is the idea of matrix big
bang\cite{Verlinde}. The authors started from IIA string theory in
a linear null dilaton background. Such a background does not
change the dimension of the spacetime and keep one-half of the
original supersymmetries. And the perturbative string theory in
the background is exactly solvable in the sense that all the
vertex operators and amplitudes could be defined formally.
However, due to the linear null dilaton the string coupling
becomes very strong near the big-bang (or big-crunch) singularity
and the perturbative string theory is ill-defined there. A dual
matrix string theory has been proposed and the physics near the
big bang is described by a weakly coupled two-dimensional
Yang-Mills theory. The matrix degrees of freedom, instead of
point-particles or strings, are used to describe the physics near
the singularity.

In \cite{Miao}, it has been pointed out that there exist other
supersymmetric backgrounds, in which one may define cosmological
matrix models. And in \cite{Das}, the matrix string theory in the
PP-wave metric with a linear null dilaton has been studied. It
would be interesting to search for other classes of time-dependent
supersymmetric configurations. Due to the existence of the
supersymmetries, the matrix models in these backgrounds could be
stable. However, it is remarkable that the standard sixteen
supersymmetries, which are characterized by the Killing spinors
satisfying $\G^+\epsilon=0$, could only be realized nonlinearly in
the matrix models and does not guarantee that the corresponding
matrix models are supersymmetric\cite{Verlinde, Das}. This
inspires us to look for the configurations with the supernumerary
supersymmetries, which may allow linearly realized supersymmetries
in their matrix models.

In section 2, we will try to find the time-dependent
supersymmetric solutions  of 11D supergravity. We obtain a large
class of configurations (\ref{metric3}), which are akin to the
usual plane-wave geometries. Such configurations keep at least
one-half of the original supersymmetries, characterized by the
Killing spinors satisfying $\G^+\epsilon=0$.

In section 3, in order to find out the extra supersymmetries and
understand the isometric symmetries, we investigate the Killing
vectors of the configurations  found in section 2. We manage to
figure out the configurations which could have extra
supersymmetries. Such configurations should have the Killing
vector with nonvanishing $\p_u$ component and being independent on
$v$. However, it turns out that all these configurations can be
transformed to the well-studied smooth homogeneous plane-waves
with supernumerary supersymmetries, including the maximal
Cahen-Wallash metric and its generalization\cite{Blau1,Blauextra}.
This leads us to conclude that in general the metric
(\ref{metric3}) can only keep the standard sixteen
supersymmetries.

 In section 4, we formulate the matrix
models in these backgrounds following the DLCQ prescription. In
particular, we focus on the backgrounds with the metrics being
exponential functions. Such backgrounds have better isometries and
are quite similar to the one studied in \cite{Das}.

\section{The 1/2 BPS configurations in 11d Supergravity}

Let us start from the equations of motions of 11-dimensional
supergravity
 \bea
 R_{MN}&=&\frac{1}{12}(F_{MPQR}F_N^{~~PQR}-\frac{1}{12}g_{MN}F^2)
 \\
 d\ast F&=& \frac{1}{2}F\wedge F
 \eea
and the Killing spinor equations
 \be
 \tilde{D}_M\epsilon=(D_M-\Omega_M)\epsilon
 \ee
 where $D_M$ is the spin connection defined by
 \be
 D_M=\p_M+\frac{1}{4}\omega^{ab}\Gamma_{ab}
 \ee
 and
 \be
 \Omega_M=\frac{1}{288}F_{PQRS}(\G^{PQRS}_{~~~~M}+8\G^{PQR}\delta^S_M).
 \ee
We would like to find the solutions to the equations of motions,
which have at least sixteen standard supersymmetries,
corresponding to the Killing spinors $\epsilon$ satisfying
$\G^+\epsilon=0$.

Let us firstly consider the solutions to the vacuum equations of
motions. Inspired by the recent studies on the time-dependent
backgrounds in string theory and M-theory\cite{Verlinde, Miao}, we
make the following ansatz:
 \be\label{metric1}
 ds^2=2A_0(u)dudv+A_i(u)(dx^i)^2,
 \ee
where the $A_0(u), A_i(u), i=1,\cdots 9$ are the functions of $u$.
For
 the metrics which are not diagonal in the $i,j$-directions but are the
functions of $u$, they could be diagonalized into (\ref{metric1}).

An orthogonal frame is
 \bea
 \th^+&=&\sqrt{A_0(u)}du \nn \\
 \th^-&=&\sqrt{A_0(u)}dv \nn \\
 \th^I&=&\sqrt{A_i(u)}dx^i\delta^I_i.
\eea The nonvanishing spin connections are
 \be
 \omega^{-i}=-\frac{\p_u\sqrt{A_i}}{\sqrt{A_0}}dx^i, \hspace{3ex}
 \omega^{-+}=-\frac{\p_u\sqrt{A_0}}{\sqrt{A_0}}du
 \ee
and the only non-zero Ricci tensor is
 \be
 R_{uu}=\sum_i\frac{\sqrt{A_0}}{\sqrt{A_i}}\left(-\p_u(\frac{\p_u\sqrt{A_i}}{\sqrt{A_0}})+\frac{\p_u\sqrt{A_i}\p_u\sqrt{A_0}}{A_0}\right).
 \ee
 Therefore we have just one equation of motion for the functions
 $A_0, A_i$'s. Next let us check the remaining supersymmetries. In
 the vacuum case, $\Omega_M=0$ and the Killing spinor equations
 are
  \bea
  \p_u\epsilon&=&-\frac{1}{2}(\omega^{+-}_u+\omega^{-+}_u\G_{-+}+\omega^{-i}_u\G_{-i})\epsilon
  \nn\\
  \p_i\epsilon&=&-\frac{1}{2}\omega^{-i}_i\G_{-k}\epsilon.
  \eea
Choosing a constant spinor $\epsilon_0$ with $\G^+\epsilon_0=0$,
then the Killing spinor satisfying the above equations is
 \be
 \epsilon(u)=\exp(-\frac{1}{2}\int\omega^{-+}_udu)\epsilon_0.
 \ee
Thus the metric (\ref{metric1}) keeps at least one-half of the
original supersymmetries.  To search for the possible extra
supersymmetries, one has to look for the Killing spinor with
$\G^+\epsilon \neq 0$.

Another class of well-studied supersymmetric configurations in 11D
supergravity is the plane-wave geometries, in the presence of a
constant 4-form field strength. Among them, the Cohen-Wallash
metric is maximal supersymmetric\cite{Papa} and the homogeneous
plane-waves have supernumerary supersymmetries. This inspires us
to look for the plane-wave-like metrics which keep at least
one-half supersymmetries, besides the metric (\ref{metric1}).

Let us introduce a constant 4-form field strength
 \be \label{F}
 F_{u123}=f_0(u),
 \ee
and make the following ansatz on the metric
 \be \label{metric2}
 ds^2=2A_0(u)dudv+B_0(u,x)du^2+A_i(u)(dx^i)^2+B_i(u,x)dx^idu,
 \ee
 where the $B_0, B_i$'s are the functions of $u$ and $x$. Temporarily we
 do not assume that the $B_0$ is quadratic in $x$ and the $B_i$'s are linear
 in $x$ so that the metric is akin to the plane-wave metric. Later on,
 we will show that the supersymmetric condition requires (\ref{Bi}) and
 then the
 equation of motion leads to (\ref{B0}).

The metric (\ref{metric2}) allows an orthogonal frame
 \bea
 \th^+&=&\sqrt{A_0(u)}du \\
 \th^-&=&\sqrt{A_0(u)}dv+\frac{B_0(u,x)}{2\sqrt{A_0(u)}}du+\frac{B_i(u,x)}{2\sqrt{A_0(u)}}dx^i
 \\
 \th^I&=&\sqrt{A_i(u)}dx^i\delta^I_i.
 \eea
The corresponding spin connections are
 \bea
 \o^{-+}&=&-\frac{\p_u\sqrt{A_0}}{\sqrt{A_0}}du \nn\\
 \o^{+i}&=&0 \nn \\
 \o^{ij}&=&-\frac{\p_iB_j}{2\sqrt{A_iA_j}}du \nn \\
 \o^{-i}&=&\frac{1}{\sqrt{A_i}}\left(\frac{\p_iB_0}{2\sqrt{A_0}}-\frac{\p_uB_i}{2\sqrt{A_0}}
 +\frac{\p_u\sqrt{A_0}}{A_0}B_i\right)du\nn\\
 & &-\frac{\p_u\sqrt{A_i}}{\sqrt{A_0}}dx^i+\sum_{j\neq i}\frac{\p_iB_j}{2\sqrt{A_0A_i}}dx^j.
  \eea

Before we give the Ricci tensor and solve the equations of
motions, let us turn to the Killing spinor equations first. With
the constant field strength (\ref{F}), we have
 \bea
 \O_v&=&0 \\
 \O_u&=&-\frac{1}{12}(\G^{+-123}+\G^{123})\frac{f_0}{\sqrt{A_1A_2A_3}}
 \\
 \O_i&=&\frac{1}{24}(3\G^{123}\G^i+\G^i\G^{123})\G^+\frac{\sqrt{A_i}f_0}{\sqrt{A_0A_1A_2A_3}},
\eea and the Killing spinor equations
 \bea
 \p_v\epsilon&=&0 \nn \\
 \p_u\epsilon&=&\O_u\epsilon-\frac{1}{2}(\o^{-+}_u+\o^{+-}_u\G^{-+}+\o^{-i}_u\G^{+i}+\o^{ij}_u\G^{ij})\epsilon\\
 \p_i\epsilon&=&\O_i\epsilon-\frac{1}{2}\o^{-k}_i\G^{+k}.\nn
 \eea
The first equation tell us that the $\epsilon$ is independent of
$v$ and if one requires it to be annihilated by $\G^+$, it is also
independent of $x^i$. In other words, introducing a $\epsilon_0$
with $\G^+\epsilon_0=0$, we have
 \be
 \epsilon(u)=\exp(x\G^{123}+y+z\G^{ij})\epsilon_0
 \ee
 with
 \be
 x=-\frac{1}{4}\int
 \frac{f_0}{\sqrt{A_1A_2A_3}}du,\hspace{3ex}y=\frac{1}{4}\ln A_0,\hspace{3ex}z=\int\frac{\p_iB_j}{4\sqrt{A_j}}du.
 \ee
And since $\p_i\epsilon=0$, one has to require $\p_iB_j$ is
independent of $x^i$. This help us to fix
 \be\label{Bi}
 B_i=A_{ij}(u)x^j
 \ee
 with $A_{ij}=-A_{ji}$\footnote{The antisymmetric condition is not necessary. We would like to thank the authors
 of \cite{Ohta} to point this out. For some more general BPS configuration, see \cite{Ohta}.}, in order to keep one-half of the
 supersymmetries.

 With the choice (\ref{Bi}), the Ricci tensor has the
 nonvanishing component
 \be
 R_{uu}=\sum_i\frac{\sqrt{A_0}}{\sqrt{A_i}}\left(-\p_u(\frac{\p_u\sqrt{A_i}}{\sqrt{A_0}})-
 \frac{1}{2\sqrt{A_0A_i}}\p^2_iB_0+
  \frac{\p_u\sqrt{A_0}\p_u\sqrt{A_i}}{A_0}+\sum_{j\neq
  i}\frac{A^2_{ij}}{4A_j\sqrt{A_0A_i}}\right).
  \ee
The nontrivial equation of motion is
 \be
 R_{uu}=\frac{f_0^2}{2A_1A_2A_3}
 \ee
The right-hand side is completely independent of $x^i$, and  in
$R_{uu}$ the only trouble comes from the term involving $B_0$. A
natural choice is to let $B_0$ be bilinear in $x^i$, namely
 \be\label{B0}
 B_0=B_{ij}(u)x^ix^j.
 \ee
In the following, we will work with (\ref{Bi},\ref{B0}). Now the
metric (\ref{metric2}) looks quite similar to the homogeneous
plane waves discussed in \cite{Blau1} but due to the dependence on
$u$, the metric (\ref{metric2}) is in general not homogeneous.

Therefore, we find a class of time-dependent supersymmetric
configurations akin to the plane-waves
 \be \label{metric3}
 ds^2=2A_0(u)dudv+B_{ij}(u)x^ix^j(du)^2+A_i(u)(dx^i)^2+A_{ij}(u)x^jdx^idu,
 \ee
 One may take
(\ref{metric1}) as a special sub-class of (\ref{metric3}).


\section{Killing vectors and Supernumerary supersymmetries}

We have shown that the configurations (\ref{metric3}) are at least
half-supersymmetric in M-theory. One interesting question is that
are there extra supersymmetries in these configurations? One may
address this issue directly by working with the Killing spinor
equations. Here we take another approach by checking if such
configurations have Killing vector with nonvanishing $\p_u$
component and no dependence on $v$. The point is that if the
Killing vector has nonvanishing $\p_u$ component, the
corresponding configurations may have extra supersymmetries. The
advantage of this approach is that it allows us to determine the
isometries of the configurations.

A Killing vector, which is of the form
 \be
 K=K^u\p_u+K^v\p_v+K^i\p_i,
 \ee
 satisfies the equation
 \be
 L_Kg_{AB}=K^C\p_Cg_{AB}+\p_AK^Cg_{CB}+\p_BK^Cg_{CA}=0.
 \ee

Without losing generality, we set $A_{ij}$ in the metric
(\ref{metric3}) to be zero. Then from the Killing equation, we
have a set of relations:
 \bea
 0&=&K^u\p_uB_0+K^i\p_iB_0+2\p_uK^uB_0+2\p_uK^vA_0 \label{pp}\\
 0&=&\p_vK^u \label{mm}\\
 0&=&K^u\p_uA_0+\p_uK^uA_0+\p_mK^mA_0 \label{pm}\\
 0&=&B_0\p_iK^u+\p_iK^vA_0+\p_uK^iA_i \label{pi}\\
 0&=&\p_vK^iA_i+\p_iK^uA_0 \label{mi} \\
 0&=&(K^u\p_uA_i)\delta_{ij}+\p_iK^jA_j+\p_jK^iA_i \label{ij}
 \eea
From (\ref{mm}), we know that $K^u$ is independent of $v$ and then
from (\ref{pm}), we find that $K^v$ is at most to be linear in
$v$. And from acting $\p_v$ on (\ref{ij}) and (\ref{mi}), we see
that the $K^u$ is at most the linear function of $x^i$, namely
 \be
 K^u=\sum_i a_i(u)x^i+a_0(u).
 \ee
Then from (\ref{pm}), we get
 \be
 K^v=-\frac{\sum_i\p_u(a_iA_0)x^i+\p_u(a_0A_0)}{A_0}v+h(u,x^i)
 \ee
 where $h(u,x^i)$ is a function to be determined. From (\ref{mi}),
 we have
 \be
 K^i=-\frac{a_iA_0}{A_i}v+e^i(u,x^i)
 \ee
 where $e^i$ is only the function of $u, x^i$. And from
 (\ref{pi}), we find that
 \bea
 a_i&=&\frac{\sqrt{c_iA_i}}{A_0} \\
 e^i&=&-\int \frac{a_iB_0+\p_ih A_0}{A_i}du +f^i(x^i).
 \eea

Next, expanding $K^i$ in powers of $x^k$
 \be
 K^i=-\frac{a_iA_0}{A_i}v+z_i(u)+z_{ik}(u)x^k+\frac{1}{2}z_{ikl}(u)x^kx^l+\frac{1}{3!}z_{iklm}(u)x^kx^lx^m
 +\cdots
 \ee
 and from (\ref{ij}), we find that the terms higher than quadratic
 vanish and $z_{ijk}=0$, if $j$ or $k\neq i$. The only
 nonvanishing coefficients are
 \be
 z_{iil}=z_{ili}=-\frac{\p_uA_i}{2A_i}a_l, \hspace{3ex}
 z_{ii}=-\frac{a_0\p_uA_i}{2A_i}
 \ee
 and $z_{ij}$'s, which satisfy the relation
 \be
 A_jz_{ji}+A_iz_{ij}=0, \hspace{5ex} \mbox{ if $i\neq j$.}
 \ee
Furthermore, taking into account of the quadratic
 form of $B_0$ and the relation (\ref{pp}), we know that $a_i$ has
 to be vanishing
 \be
 a_i=0.
 \ee
 From the linear term in $v$, quadratic terms $vx^j$ in
  (\ref{pp}), we get the following relations
  \bea
  a_0(u)&=&\alpha\frac{\int A_0du}{A_0}+\beta, \label{a0}\\
  \p_uh_0(u)&=&0, \\
  h_{ii}(u)&=&\frac{A_i}{A_0}\p_u(\frac{a_0\p_uA_i}{2A_i}),
  \label{gii} \\
  \p_uh_j&=&\sum_i\frac{B_{ji}}{A_0}\int \frac{h_iA_0}{A_i}du,
  \label{gi}
  \eea
  where $\alpha,\beta$ are constants and  $h_0,h_i,h_{ii}$ are the expansive coefficients of the function $h$ in terms of $x$:
  \be
  h(u,x^i)=h_0(u)+h_i(u)x^i+\frac{1}{2}h_{ij}x^ix^j+ \cdots.
  \ee
And finally from the quadratic $x^jx^k$ terms in (\ref{pp}),
  we get
  \bea
  0&=&a_0\p_uB_{ii}-\frac{a_0\p_uA_i}{A_i}B_{ii}+2\p_ua_0B_{ii}+2\sum_jf_{ji}B_{ji}+2\p_uh_{ii}A_0,
  \label{Aii} \\
  0&=&a_0\p_uB_{jk}+2\p_ua_0B_{jk}-\frac{a_0\p_uA_j}{A_j}B_{jk}+2\sum_if_{ik}B_{ij}, \hspace{3ex} j\neq k \label{Ajk}
  \eea
with the constants $f_{ij}$'s constrained by
 \be \label{fij}
 A_jf_{ji}+A_if_{ij}=0.
 \ee Then the
components of the Killing vector read
 \bea
 K^u&=&a_0(u), \\
 K^v&=&-\frac{\p_u(a_0A_0)}{A_0}v+h_0(u)+h_i(u)x^i+h_{ii}(u)(x^i)^2,
 \\
 K^i&=&-\int
 \frac{h_iA_0}{A_i}du-\frac{a_0\p_uA_i}{2A_i}x^i+f_{ik}x^k
 \eea

From the constraint (\ref{fij}) on the constants $f_{ij}$'s, we
know that if $A_i(u)$ is not proportional to $A_j(u)$, $f_{ij}=0$.
This indicates that the usual rotational Killing symmetry
$x^i\p_j-x^j\p_i$ gets lost in this case.

Since either $f_{ij}$ being antisymmetric or being zero, the
relevant term $\sum_jf_{ji}B_{ji}$ in (\ref{Aii}) is vanishing and
(\ref{Aii}) gives us a relation between $a_0$ and $B_{ii}$ once
the form of the metric is fixed. In order to have nonvanishing
$K^u$, the form of the metric is highly constrained by
(\ref{Aii},\ref{Ajk}).

The cases when the Killing vector has nonvanishing $K^u$ component
are interesting since the corresponding configurations may have
extra supersymmetries. This could be seen as
follows\cite{Blauextra}. Consider the Killing vector
 \be
 K=\bar{\epsilon}\G^\mu\epsilon\p_\mu,
 \ee
which has the component
$K^u=\frac{1}{\sqrt{2}}(\G^+\epsilon)^T(\G^+\epsilon)$. For the
supersymmetric configurations we discussed above, we always have
sixteen standard supersymmetries, characterized by the Killing
spinors satisfying $\G^+\epsilon=0$. These Killing spinors can not
give us  the nonvanishing $K^u$. However, the extra
supersymmetries with $\G^+\epsilon \neq 0$ will give us the
nonvanishing Killing component $K^u$. On the other hand, the
metric with a nonvanishing Killing vector $K^u$ could have
supernumerary supersymmetries.

Let us figure out in which case we can have the Killing vector
with a nonvanishing $K^u$. Firstly let us consider the case when
$B_0=0$. From (\ref{Aii}) we know that $g_{ii}$ should be
constant. Only in very special situation, say all $A_i\propto A_0$
and $A_i$'s being the exponential function of $u^2$, one has a
nonvanishing constant $g_{ii}$. Generically $g_{ii}=0$, which
require
 \begin{itemize}
 \item $a_0=0$;
 \item $A_i$'s being constant with no restriction on
$a_0$;
 \item $A_i$'s being exponential function with $a_0$ being
constant.
 \end{itemize}
 Moreover, from $\p_uh_j=0$ we get
$h_j=\alpha_ju+\beta_j$ with $\alpha_j,\beta_j$ being constant.

Let us see a few examples:
\begin{itemize}
\item The trivial example is that all the $A_0, A_i$'s are
constant, which could be normalized to be unit. Obviously, all the
discussions above reduce to the study of the Killing vectors in
the flat spacetime;

\item The first nontrivial example is the configuration
corresponding to the null linear dilaton
background\cite{Verlinde}:
 \be
 ds^2=e^{2Qu/3}ds_{10}^2+e^{-4Qu/3}(dx^{11})^2
 \ee
 where $Q$ is a constant. In this case, the requirement that
 $h_{99}$ is constant lead to $a_0=\beta$ and $h_{ii}=0$. And
 since $A_i=A_0$, for $i=1,\cdots 8$, one has rotational symmetry
 among $x^i$'s and also a rotational symmetry between $u$ and
 $x^i$.

\item A more general case with vanishing $B_0$ is that all the
$A_i(u)$'s are exponential functions but different so that
$f_{ij}=0$. The components of the Killing vectors could be of the
form:
 \bea
 K^u&=&\beta \nn\\
 K^v&=&-\beta(\p_u\ln A_0)v+h_0(u)+\sum_ih_i(u)x^i \nn\\
 K^i&=&-\int \frac{h_iA_0}{A_i}du-\frac{\beta}{2}(\p_u\ln A_i)x^i
 \eea

 \end{itemize}

 Next we turn to the more general metrics with a nonvanishing $B_0$.
 From (\ref{Aii}), we know that it is hard to find nontrivial
 solution when $h_{ii}\neq 0$ and generically we have $a_0=0$. A
 case with a nonvanishing $a_0$ is that when $A_i(u)$ take the
 exponential forms $A_i(u)=\exp(\gamma_i u)$ and $a_0=\beta$,
  \be
  B_{ii}=c_i A_i
  \ee
  with $c_i$ being constants. One may set $\beta=1$ and then
  \be \label{B01}
  B_{jk}=e^{\gamma_j u}(e^{2uf}B_se^{-2uf})_{jk}
  \ee
  where $B_s$ is a constant symmetric matrix. It is always
  possible to arrange the matrix $f_{ij}$ so that the matrix $B_{jk}$ takes a block form. In
  other words, if $A_j\neq A_k$, $B_{jk}=0$.

  Another special case
  is when all $A_i$ are constants and then one may has $a_0=u$,
  and
   \be \label{B02}
   B=\frac{1}{u^2}e^{2f\log u}B_se^{-2f\log u}.
   \ee

 Usually, a Killing vector with nonvanishing $K^u$ component is not enough to ensure the
 extra supersymmetries. Since in our discussion the Killing
spinor is independent of $v$, the corresponding Killing vector
cannot have $v$ dependence. From the form of $K^v$, the condition
to have no term proportional to $v$ is that $A_0=\mbox{constant}$
which could be set to 1.

 From the
above discussion we learn that the nonvanishing $K^u$ gives the
strong constraint on the form of the metric. In the case with
$B_0=0$, we know that when $A_0, A_i$ are exponential functions,
we have $K^u=\beta$ which could be chosen to be $1$ for
simplicity. And one need to take $A_0=1$ so that $K^v=0$. But we
always has a nonvanishing term $-\frac{\gamma_i}{2}x^i$ in $K^i$
so the Killing vector is
 \be
 K=\p_u-\frac{\gamma_i}{2}x^i\p_i.
 \ee
 after choosing $h_i$'s vanishing.
In the cases discussed in \cite{Verlinde, Miao}, $A_0, A_i$'s take
the exponential forms, but since $A_0\neq 1$, the background
cannot have the supernumerary supersymmetry.

In the case with nonvanishing $B_0$, the existence of the
nonvanishing $K^u$ requires $B_{jk}$ take the form (\ref{B01}) or
(\ref{B02}). In the latter case, the existence of nonvanishing
$K^v$ shows that there is no extra supersymmetry. On the other
hand, in the case (\ref{B01}), there may exist the extra
supersymmetries when $A_0=1$.

At this moment, let us do a short summary. The time-dependent
supersymmetric configurations with the possible supernumerary
supersymmetries are
 \begin{itemize}
 \item $B_0=0, A_0=1$, and $A_i$'s are exponential functions;
 \item $B_0 \neq 0, A_0=1$, $A_i$'s are exponential functions and $B_{jk}$ take the form of
 (\ref{B01}).
 \end{itemize}
So let us focus on the metrics of the form \be\label{me3}
 ds^2=2dudv+\sum_ie^{\gamma_i
 u}(dx^i)^2+\sum_{ij}B_{ij}x^ix^j(du)^2.
 \ee
 However, it is easy to see that the above
metrics could be transformed to the well-studied ones by changing
coordinates. In fact, after defining
 \bea
 \tilde{x^i}&=&e^{\g_iu/2}x^i \nn \\
 \tilde{v}&=&v-\sum_i\frac{\g_i}{4}e^{\g_i u}(x^i)^2,
 \eea
the above metric takes a form
 \be
 ds^2=2dud\tilde{v}+(d\tilde{x}^i)^2+\sum_i\frac{\g_i^2}{4}(\tilde{x}^i)^2(du)^2
 +\sum_{ij}\tilde{B}_{ij}\tilde{x}^i\tilde{x}^j(du)^2
 \ee
 where
  \be
 \tilde{B}_{ij}=(e^{2uf}B_se^{-2uf})_{ij}.
  \ee
On the other hand, in the new coordinates $(u, \tilde{v},
\tilde{x}^i)$, the 4-form field strength is
 \be
 F_{u\tilde{1}\tilde{2}\tilde{3}}=\mbox{constant}.
 \ee
Therefore, in terms of the new coordinates  the background
(\ref{me3}) with the possible supernumerary supersymmetries  is
reduced to the well-studied  homogeneous
plane-waves\cite{Blauextra}.

We conclude that our
 time-dependent supersymmetric configurations of the metric form
 (\ref{metric3}) have no supernumerary
 supersymmetry, except the cases with constant $A_0, A_i$'s and
 the appropriately chosen $B_{ij}, A_{ij}$'s.

\section{Matrix models}

In general, the definition of the matrix models in the curved
backgrounds and the time-dependent backgrounds is a subtle issue.
In our case, we are not be able to derive the matrix models
following the argument in \cite{Sen, Seiberg}. Instead, we obtain
the matrix models by considering the lightcone gauge action of a
single massless particle in eleven dimensions with momentum
$p_-=N/R$\cite{Taylor}. The bosonic action of the matrix model in
a curved background is
 \bea \label{matrixB}
 S_B&=&\int d\tau
 STr\left(\frac{1}{2R}(g^{uv})g_{ij}D_\tau X^iD_\tau X^j-\frac{1}{R}g_{ij}g^{vi}D_\tau
  X^j+\frac{1}{2g^{uv}R}(g_{ij}g^{vi}g^{vj}-g^{vv}) \right.\nn\\
 & & \left. +\frac{R}{4g^{uv}}g_{ik}g_{jl}[X^i,X^j][X^k,
 X^l]-\frac{i}{2}A_{uij}X^iX^j\right)
 \eea
 and the fermionic action is of the form
 \be \label{matrixF}
 S_F=\int d\tau STr \left(i{\psi}^TD_\tau \psi- \frac{1}{R}g_{ij}\psi^T\G^i[X^j,
 \psi]+\frac{i}{48R}\psi^T \G^{ijk}\psi
 F_{uijk}+\frac{1}{16R}\psi^T\G^{ij}\psi \p_j(g_{+i})\right)
 \ee
Note that in the above relation we use a loose notation where the
index in $\G$ is not in the frame.

We have seen that the metric (\ref{metric3}) with $A_0,B_0,A_i$'s
being exponential have better geometric property from the
discussion on the Killing vector. Let
 \be \label{exp1}
 A_0=e^{\g_0 u}, \hspace{3ex} A_i=e^{\g_i u}, \hspace{3ex}
 B_0=\sum_i c_ie^{\g_i u}(x^i)^2,
 \ee
 where $c_i$'s being constant, and for simplicity, we let $B_0$ to be
 diagonal.
From the equation of motion, we have
 \be \label{exp2}
 A_{ij}=A_{ij}^0e^{(\g_i+\g_j)u/2}, \hspace{3ex}
 F_{uijk}=e^{\frac{(\g_i+\g_j+\g_k)}{2}u}\epsilon_{ijk}f^0
 \ee
 where $A_{ij}^0=-A_{ji}^0=\mbox{constant}$, $f^0$ being constant.

With respect to the metric (\ref{metric3}) and (\ref{exp1},
\ref{exp2}), the action of the matrix models is
 \bea
 S_B&=&\int d\tau
 STr\left(\frac{1}{2R}\sum_i e^{(\g_i-\g_0)\tau}(D_\tau
 X^i)^2+\frac{1}{2R}\sum_{ij}e^{(\frac{\g_i+\g_j}{2}-\g_0)\tau}A^0_{ij}X^jD_\tau
 X^i+\frac{1}{2R}\sum_ie^{(\g_i-\g_0)\tau}c_i(X^i)^2 \right.\nn\\
 & &\left. +\frac{R}{4}\sum_{ij}e^{(\g_0+\g_i+\g_j)\tau}[X^i,
 X^j]^2+\frac{if^0}{2}e^{\frac{(\g_i+\g_j+\g_k)}{2}\tau}\epsilon_{ijk}X^iX^jX^k
 \right) \nn\\
 S_F&=&\int d\tau STr \left(i{\psi}^TD_\tau
 \psi-\frac{1}{R}\sum_i e^{\frac{\g_i}{2}\tau}\psi^T\G^i[X^i,\psi]+\frac{if^0}{8R}\psi^T\G^{ijk}\psi
 +\sum_{ij}\frac{A^0_{ij}}{16R}\psi^T\G^{ij}\psi\right) \label{matrixexp}
 \eea

Although we have formally defined the matrix models in the
time-dependent supersymmetric backgrounds, we are not certain if
such a definition make sense or not. The point is that the action
of the matrix model  (\ref{matrixB}, \ref{matrixF}) is defined in
a {\it weakly} curved background, which require the background
deviate from the flat spacetime not very much. However, in our
cases, the metric dependence on $u$ is of exponential form such
that the background could be far from flat. Especially, near the
big-bang or big-crunch singularity, the metric could be singular.
One may needs to check the consistency of the action. In
\cite{Das}, a matrix string theory action in a null linear dilaton
PP-wave background has been proposed and was argued to truly
describe strongly coupled string theory. Due to the close relation
between the matrix string and the matrix model and the similarity
between our background and the one in \cite{Das}, we are inclined
to believe that the action (\ref{matrixexp}) make sense.

It would be interesting to study the physical properties of these
matrix models and investigate their possible cosmological
applications. Also it could be important to check the stability of
these backgrounds.

\section*{Acknowledgement}

We would like to thank Miao Li for reading the manuscript and
valuable suggestions. We are grateful to Interdisciplinary Center
for Theoretical Studies, where the work was finished. The work was
supported by CNSF Grant No. 10405028 and the Key Grant Project of
Chinese Ministry of Education (NO. 305001).

\end{document}